\documentclass[twocolumn]{aastex61}
\usepackage{amsmath,amstext}
\usepackage[T1]{fontenc}
\usepackage{apjfonts} 
\usepackage[figure,figure*]{hypcap}
\makeatletter

\newcommand{\Rmnum}[1]{\expandafter\@slowromancap\romannumeral #1@}
\makeatother

\usepackage{color}

\usepackage{soul}
\usepackage[colorinlistoftodos]{todonotes}

\shorttitle{The Case of Cetus II}
\shortauthors{Conn et al.}

\begin{document}

\title{On The Nature of ultra-faint Dwarf Galaxy Candidates II: The case of Cetus II}

\author{Blair C. Conn}
\affiliation{Research School of Astronomy and Astrophysics, Australian National University, Canberra, ACT 2611, Australia}

\author{Helmut Jerjen}
\affiliation{Research School of Astronomy and Astrophysics, Australian National University, Canberra, ACT 2611, Australia}

\author{Dongwon Kim} 
\affiliation{Dept. of Astronomy, University of California, Berkeley, CA 94720}

\author{Mischa Schirmer} 
\affiliation{Max-Planck-Institut f\"ur Astronomie, K\"onigstuhl 17, 69117 Heidelberg, Germany}
\affiliation{Gemini Observatory, Casilla 603, La Serena, Chile}

\correspondingauthor{Blair C. Conn}
\email{blair.conn@anu.edu.au}

\begin{abstract} 
We obtained deep Gemini GMOS-S $g,r$ photometry of the ultra-faint dwarf galaxy candidate Cetus II with the aim of providing stronger constraints on its size, luminosity and stellar population. Cetus II is an important object in the size-luminosity plane as it occupies the transition zone between dwarf galaxies and star clusters. All known objects smaller than Cetus II ($r_h \sim 20$\,pc) are reported to be star clusters, while most larger objects are likely dwarf galaxies. We found a prominent excess of main-sequence stars in the colour-magnitude diagram of Cetus II, best described by a single stellar population with an age of 11.2\,Gyr, metallicity of [Fe/H] = $-1.28$\,dex, an [$\alpha$/Fe] = 0.0\,dex at a heliocentric distance of 26.3$\pm$1.2\,kpc. As well as being spatially located within the Sagittarius dwarf tidal stream, these properties are well matched to the Sagittarius galaxy's Population\,B stars. Interestingly, like our recent findings on the ultra-faint dwarf galaxy candidate Tucana V, the stellar field in the direction of Cetus II shows no evidence of a concentrated overdensity despite tracing the main sequence for over six magnitudes. These results strongly support the picture that Cetus II is not an ultra-faint stellar system in the Milky Way halo, but made up of stars from the Sagittarius tidal stream.

\end{abstract}

\keywords{
Stars: Hertzsprung-Russell and C-M diagrams -- 
Galaxy: globular clusters: general -
Galaxy: halo --
galaxies: dwarf -- 
galaxies: individual (Sagittarius) --
Local Group}

\section{Introduction}
In Paper\,I \citep{Conn2018} we demonstrated that the ultra-faint dwarf galaxy candidate Tucana V, also known as DES J2337-6316~\citep{Drlica-Wagner2015} does not have the stellar concentration typical for an ultra-faint star cluster or dwarf galaxy. 

Our results based on deep stellar photometry led to the conclusion that Tucana V, originally detected at a significance level of $\sigma=8.0$, must be either the debris of a completely tidally disrupted star cluster or an excess of stars in the halo of the Small Magellanic Cloud. In regard to the search for ultra-faint stellar systems in the Milky Way halo, we propose that Tucana V is an example of a false-positive detection. This raises concern that other candidates reported in the literature and taken at face value by other researchers are in fact false-positives too. We highlighted the region around Tucana V in the size-luminosity plane as a ``Trough of Uncertainty" regarding these types of objects. The other two known objects which reside in that region are Draco II \citep{Laevens2015b} and Cetus II \citep[DESJ0117-1725]{Drlica-Wagner2015}. Cetus II is the focus of this paper.

Cetus II has a reported heliocentric distance of $d_\odot=30\pm$3\,kpc, a half-light radius $r_h = {1.9}_{-0.5}^{+1.0}$ arcmin and a total luminosity of $M_V = 0.00\pm0.68$ \citep{Drlica-Wagner2015}.
It also has the lowest detection significance ($\sigma=5.5$) of all objects reported by \citet{Drlica-Wagner2015} as estimated from their stellar density map search method.
These authors further noted that Cetus II should be treated with caution due to inter-CCD gaps in the DES\footnote{Dark Energy Survey, http://des.ncsa.illinois.edu/releases/sva1D} data available at that time. However, if confirmed it would be the least luminous galaxy known to date. In this paper we seek to better understand the phenomenon  Cetus II and refine the object's properties by obtaining deep photometry with the GMOS-S instrument. We also want to determine whether its location in the Trough of Uncertainty unveils it as another false-positive detection or a true ultra-faint dwarf galaxy candidate.

The rapid increase in the number of known Milky Way satellites over the last couple of years \citep{Balbinot2013, Belokurov2014,Laevens2014, Bechtol2015, Drlica-Wagner2015, Kim2, Kim2015b, KimJerjen2015a, KimJerjen2015b, Koposov2015, Laevens2015a, Laevens2015b, Martin2015, Kim2016, Luque2016, Martin2016b, Torrealba2016a, Torrealba2016b, Koposov2017} has important implications to our understanding of galaxy formation and near-field cosmology. In particular, the newest discoveries are some of the smallest bound stellar systems and thus constitute prime laboratories to study star formation on the smallest scales, in pure baryonic and dark matter dominated environments. At the ultra-faint end of the satellite galaxy luminosity function, there are still relatively few objects, which receive a high statistical weight in studies that correct observed satellite counts for detection efficiency. Consequently, any misclassification of a ultra-faint dwarf galaxy can skew the results significantly. Hence, it is imperative to know the true nature of every single object.

In $\S$\ref{sec:observations} we present the details of our follow-up observations of Cetus II, the photometric calibration procedure, the artificial star experiment and the colour-magnitude diagram for all the stars we detected in the Cetus II field. In $\S$\ref{sec:CetIIpop} we revisit the adopted procedure for determining the age, metallicity and distance of the Cetus II stellar population and present the results. In $\S$\ref{sec:discussion} we discuss our findings and draw conclusions about the nature of Cetus II in $\S$\ref{sec:conclusion}.

\begin{figure*}
\begin{center}
\includegraphics[width=1.0\hsize]{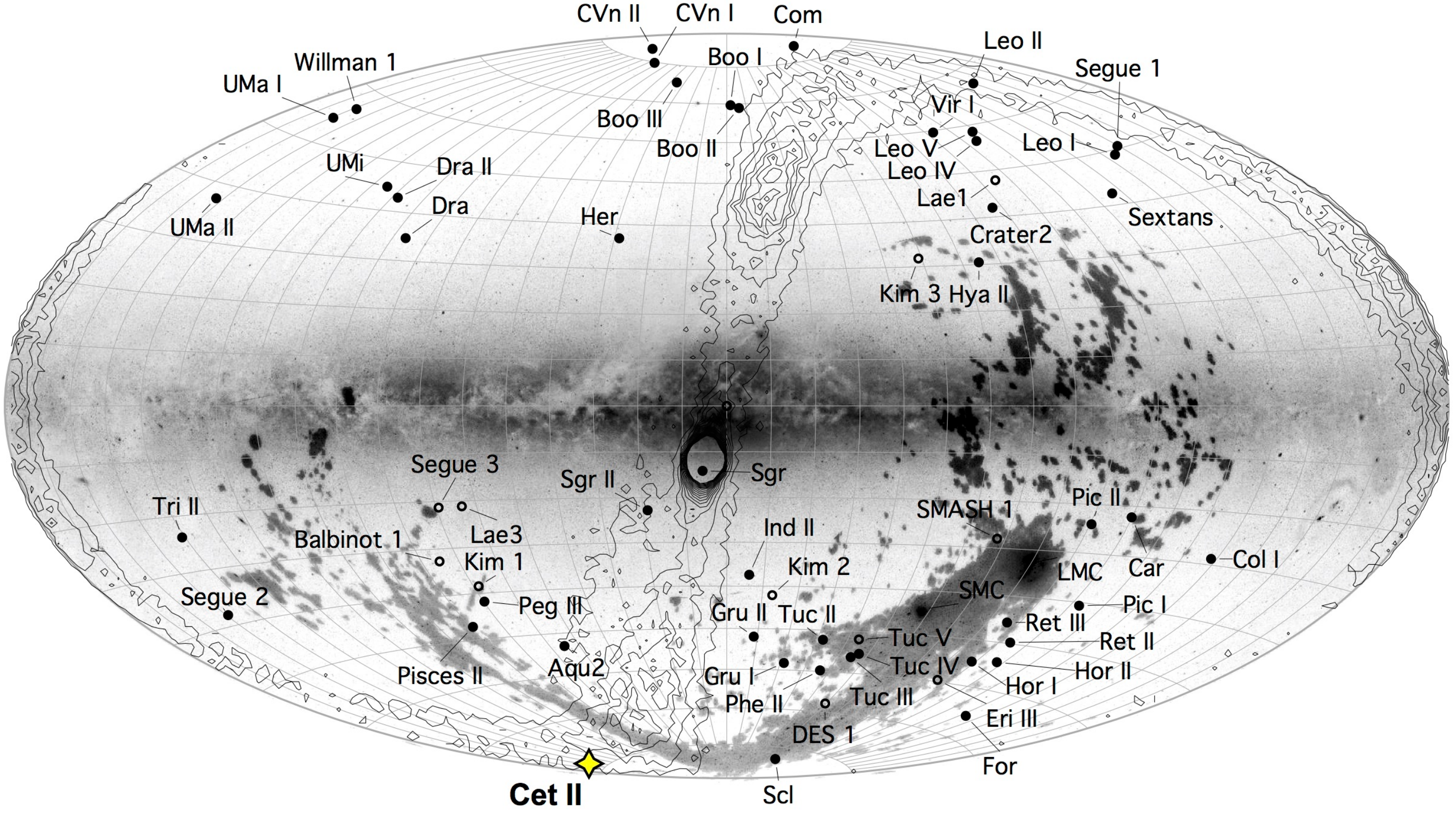}
\caption{All-sky view in Galactic coordinates showing the distribution of known Milky Way satellite dwarf galaxies 
(filled circles) and star clusters (open circles). Cetus II (yellow dot) is found close to the Galactic South pole at: $l=156\fdg47$, $b=-78\fdg53$, superimposed on the Sagittarius stellar tidal stream (contours) and close to the neutral hydrogen gas of the Magellanic Stream (grey scale image). The star density contours of the Sagittarius stream are inferred from the \citet{LM2010} tidal debris model, which adopts a triaxial dark matter halo for the Milky Way.}\label{fig:MWS}
\end{center}
\end{figure*}

\section{Observations and Data Reduction}\label{sec:observations}
\begin{figure}
\begin{center}
\includegraphics[width=1.0\hsize]{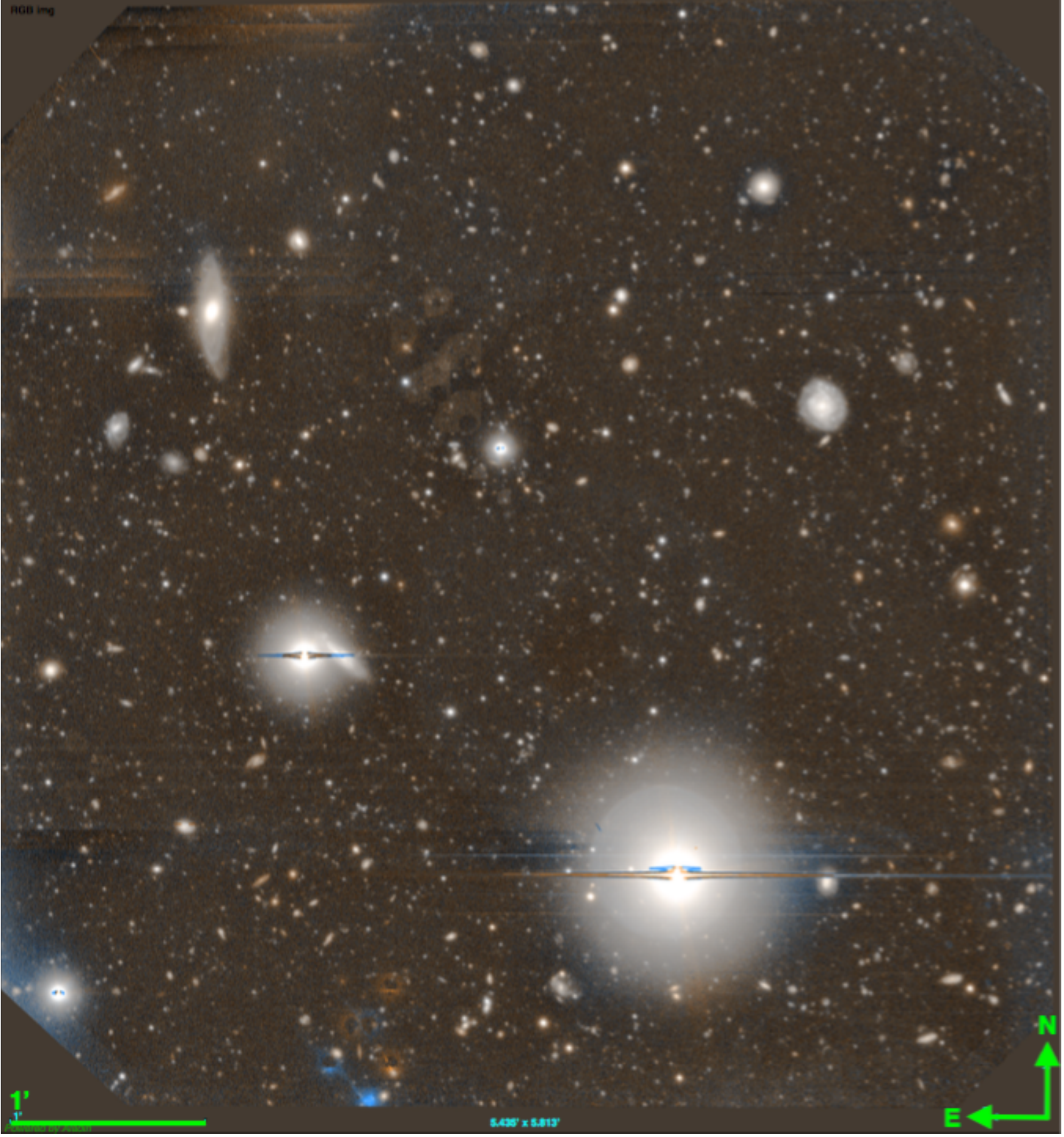}
\caption{False colour RGB image of the GMOS-S field centred on Cetus II, made using {\sc Aladin Sky Atlas v8.040}. The $g$ and $r$-band co-added images were used for the blue and red channel, respectively. Owing to the exquisite seeing, a large number of background galaxies are recognizable. However, no stellar overdensity is visible in the field. The bar in the lower left corner has a length of 1 arcminute.}
\label{fig:CetII}
\end{center}
\end{figure}

\begin{table*}
\caption{Observations }
\centering
\begin{tabular}{lrccccccc}
\hline
Field & Right Ascension & Declination & Position Angle & Filter & Observation & Airmass & Exposure & Seeing\\
 &  (deg, J2000) & (deg, J2000) & (deg) & & Date& & (sec) & (\arcsec)\\
 \hline\hline
Cetus II & 19.4667& $-$17.425& 90&g\_G0325 & 2017-09-17& 1.04 - 1.06& $1\times 60$, $3\times 600$ & 0.60\\
(DES J0117-1725) &   &           &    90 &   r\_G0326   &      2017-09-17  &    1.07 - 1.12 &  $1\times$ 60, $3\times$ 600  &   0.54    \\
\hline
\end{tabular}
\end{table*}\label{table:data}

The imaging data presented here were obtained with the Gemini Multi-Object Spectrograph South (GMOS-S) at the 8m-class Gemini South Observatory through Program ID: GS-2017B-Q-40.
The observing conditions required for the observations to be scheduled, following the Gemini Observatory standards, were dark\footnote{SB50 - Sky Brightness 50$^{th}$ percentile}, clear skies\footnote{CC50 - Cloud Cover 50$^{th}$ percentile and excellent seeing\footnote{IQ20 - Image Quality 20$^{th}$ percentile}}. On average we achieved $0\farcs60$ in the $g-$band (g\_G0325) and $0\farcs54$ in the $r-$band (r\_G0326) for the night of September 17, 2017 (see Table~\ref{table:data}). The superb seeing obtained in IQ20 conditions allowed us to utilize the $1\times1$ binning mode of GMOS-S, which affords a pixel scale of $0\farcs08$. The GMOS-S field of view is $5\farcm5\times 5\farcm5$ and our observing strategy involved having a short 60\,sec exposure centred on the target and three dithered exposures of 600\,sec each. We employed the {\sc theli} pipeline \citep{2013ApJS..209...21S} to perform the basic data reduction of creating a master bias and master twilight flats, bias subtraction and flat fielding, astrometry and co-addition. To generate the object catalogues we used the Point Spread Function (PSF) photometry package {\sc dolphot} \citep{2000PASP..112.1383D} on the combined stacked images.

{\sc Dolphot} has approximately 80 parameters which need to be set to process the data, the majority of these relate to the fundamental details of the files being processed: filenames, filters, offsets between frames in pixels, initial estimates of the seeing in pixels, exposure time, read noise, bad pixel value, saturation value, airmass etc. We chose to use a point spread function based on a linear Gaussian + Lorentzian solution and we set the detection threshold to 2.5 sigma above the noise.

\subsection{Photometric Calibration}\label{sec:calibration}

\begin{table}
\caption{Photometric Calibration Results}
\begin{center}
\begin{tabular}{lcc}
\hline
 & $g$ band & $r$ band \\ \hline\hline 
Colour term $(g -r)$\footnote{from \citet{Conn2018}} & $+0.026^{+0.045}_{-0.046}$& $-0.059^{+0.042}_{-0.041}$\\
Cetus II offsets& $-3.018^{+0.028}_{-0.029}$& $-2.771^{+0.026}_{-0.027}$ \\\hline
\end{tabular}
\end{center}
\tablecomments{Colour terms and offsets derived from comparison with APASS calibrated DES photometry. All photometry assumed a zeropoint of 30.00 for both filters prior to the offsets being applied.The offset values listed are a combination of the true zeropoint correction and the atmospheric extinction correction.}\label{table:calib}
\end{table}

For the calibration of the instrumental magnitudes generated by the {\sc dolphot} photometry, we followed  the same steps as discussed in Paper I. The GMOS-S data were cross-matched with APASS\footnote{The AAVSO Photometric All-Sky Survey}~\citep{2015AAS...22533616H} calibrated DECam photometry\footnote{DECam photometry generated using the procedures outlined in \citet{KimJerjen2015b}}. We utilized the same technique as described in Paper I, applying the same colour term and then calculating the linear offset (a combination of a zeropoint offset and atmospheric extinction correction) from a generic instrumental zeropoint. The calibration proceeds by first applying the offset directly to the raw magnitudes and then iterating the magnitudes using the colour and colour term to achieve the final solution, as shown in Equation~\ref{eqn:calib}.
Apply the offset:
\begin{eqnarray}\label{eqn:calib}
\text{rawmag} = \text{rawmag} + \text{offset}
\end{eqnarray}
Iterate the magnitudes using:
\begin{eqnarray}\label{eqn:calib2}
\text{newmag} = \text{rawmag} + \text{Colour Term}*(g-r)
\end{eqnarray}
After each iteration update {\it rawmag} to the new {\it newmag} value then repeat until the solution converges. These colour terms and offsets, applied to the Cetus II field, are listed in Table~\ref{table:calib}. The selection criteria for objects to be included in the final catalogue consisted of finding objects where:
\begin{itemize} 
	\item in either filter, sharpness$^{2} \leq 0.1$
	\item in both filters, signal-to-noise ratio $\geq3.5$
    \item and the object type corresponds to "good stars" (Objtype = 1). 
\end{itemize}
Spurious or saturated objects were removed from the catalogue through either their extremely large magnitude errors or zero magnitude error respectively.

\subsection{Artificial Star Experiments}\label{sec:artstars}
\begin{figure}
	\centering
	\includegraphics[width=1.0\hsize]{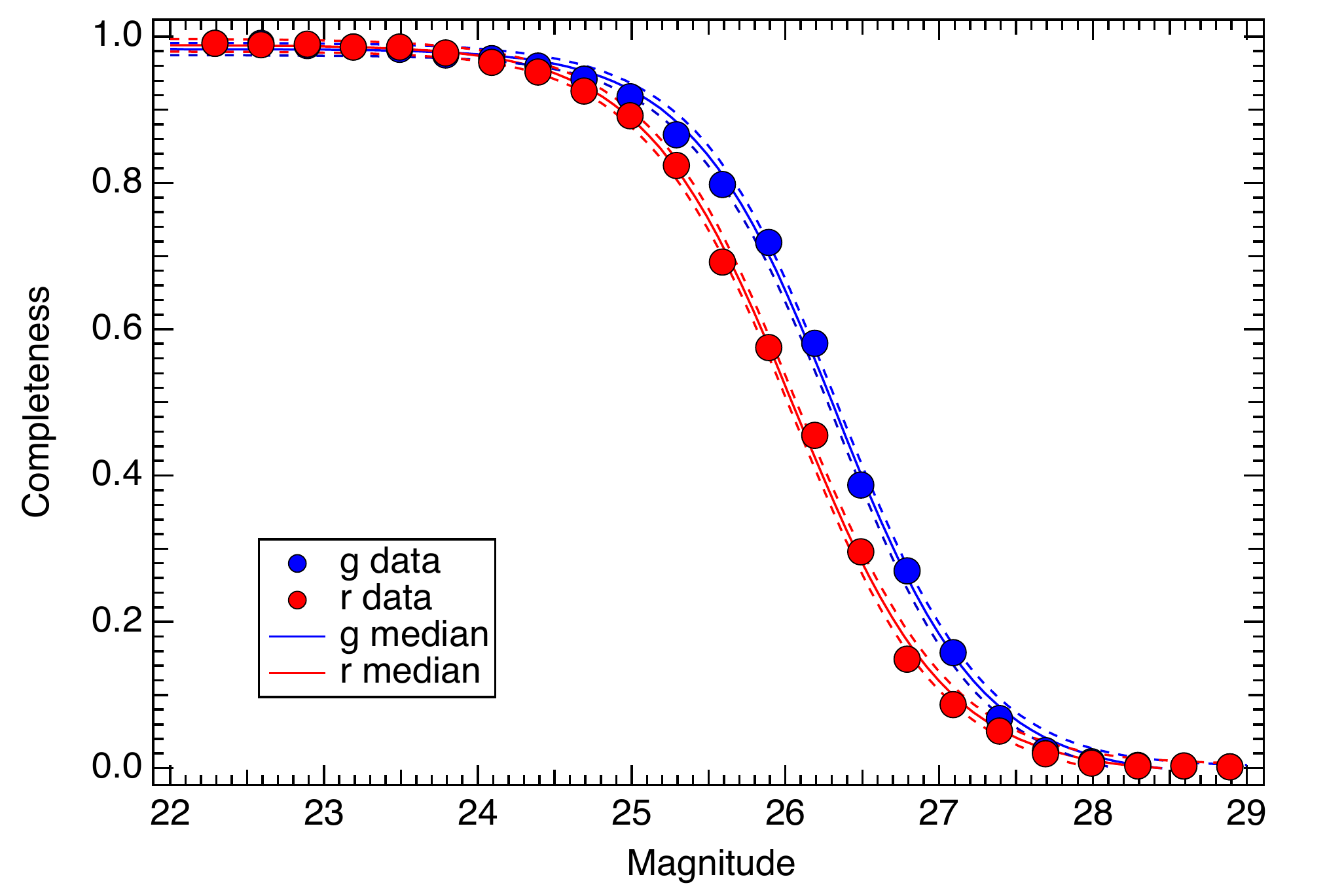}
	\caption{\label{fig:completeness} Recovery rate for artificial stars in the Cetus II field. The points show the photometric completeness per 0.3 magnitude bin, while the solid line shows the best fit solution with the dashed lines highlighting the 90th percentiles.}

\end{figure}

Following the procedure described in Paper I, the photometric completeness of the Cetus II field was determined using the {\sc dolphot} built-in artificial star experiment by generating a flat luminosity function with around 500,000 stars covering the magnitude interval $20<m<29$ and subdivided in 0.3\,mag bins. The recovery rate in each filter was fitted with a Logistic function: 
\begin{eqnarray}\label{eqn:logistic}
Completeness = (1 + e^{(m - mc)/\lambda})^{-1}
\end{eqnarray}
where $m$ is the magnitude, $mc$ is the 50\% completeness value and $\lambda$ is the width of the rollover.  The parameters of the best-fit solutions are listed in Table~\ref{table:completeness} and Figure~\ref{fig:completeness}.  

\begin{figure*}
\begin{center}
	\includegraphics[width=1.0\hsize]{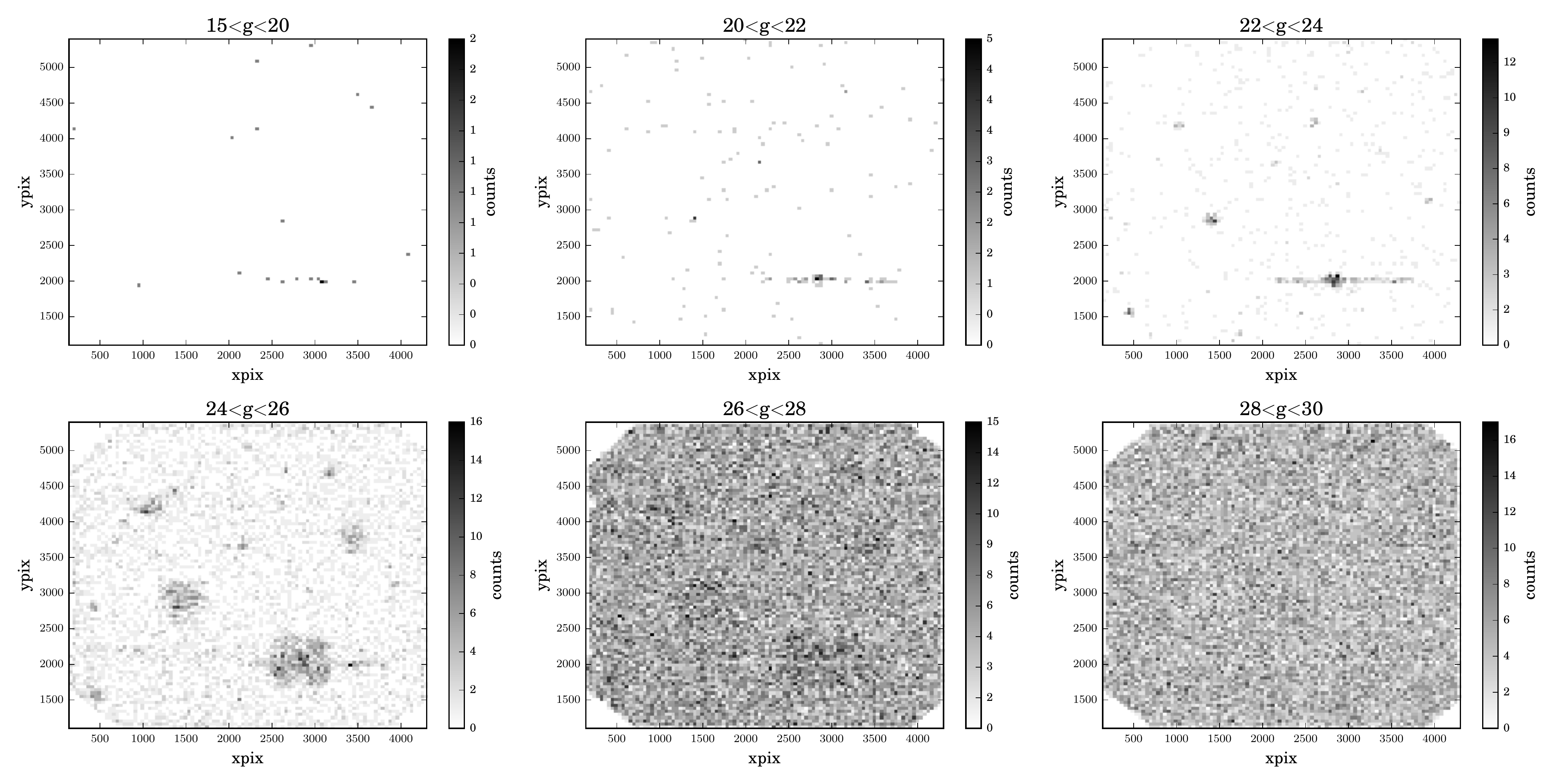}
	\caption{\label{fig:loststars} Pixel positions of unrecovered artificial stars sorted by $g$-band magnitude.}
\end{center}
\end{figure*}

To explore the impact of the bright stars in the field on both the photometric completeness and any potential overdensity in the field, we show the position of the unrecovered artificial stars sorted by $g$-band magnitude in Figure~\ref{fig:loststars}. Here we can see clearly the effect of the bright stars and their halos in the field. Interestingly the biggest effect is in the magnitude range $24<g<26$, whereas at fainter magnitudes (greater than the 50\% completeness level) the distribution of unrecovered stars becomes much smoother. In general though, the amount of the field lost to the bright stars is quite small as can be seen in Figure~\ref{fig:completeness} where, at the brighter magnitudes, the completeness is very close to 100\%. Significant loss of coverage would lower the maximum completeness level proportional to the amount of contamination.

\begin{table}
\caption{50\% Photometric Completeness Estimates}\label{table:completeness}
\centering
\begin{tabular}{cccc}
\hline
 $mc_g$ & $\lambda_g$&$mc_r$& $\lambda_r$ \\\hline\hline
26.33$\pm{0.04}$ & 0.473$\pm{0.030}$ & 26.06$\pm{0.03}$ & 0.486$\pm{0.030}$ \\
\hline
\end{tabular}

\end{table}

\subsection{Colour-Magnitude Diagram}\label{sec:cmd}
\begin{figure}
\begin{center} 
\includegraphics[width=1\hsize]{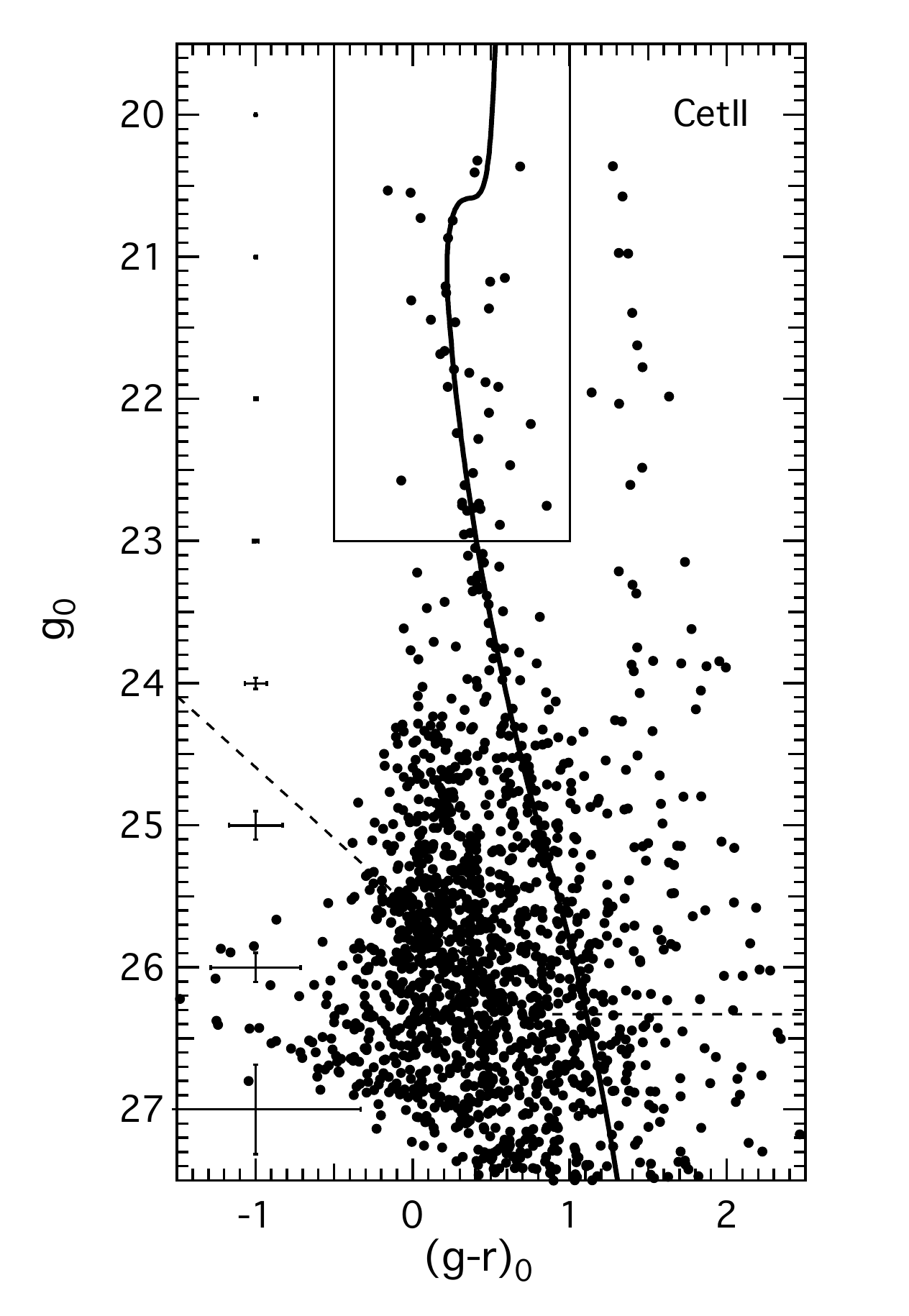}
\caption{The $g_\circ$ vs. $(g-r)_\circ$ colour-magnitude diagram of objects classified as point sources in the $5\farcm5\times 5\farcm5$ GMOS-S field centred on Cetus II. The rectangular window corresponds to the colour-magnitude parameter space investigated in the discovery paper \citep[Fig.\,14 in][]{Drlica-Wagner2015}. The CMD reveals a distinct main sequence population extending over six magnitudes down to $g\sim 26.8$. The error bars running vertically along the colour axis in 1\,mag intervals represent the typical photometric uncertainties. The 50\% completeness level can be seen as a dashed line. \label{fig:cmd_field}}
\end{center}
\end{figure}

Figure~\ref{fig:cmd_field} shows the extinction-corrected $(g-r)_\circ, g_\circ$ CMD of the  GMOS-S field using all point sources from our photometry analysis found in the vicinity of Cetus II. The window in the brighter section of the CMD highlights the region that was investigated in the discovery paper derived from DES data \citep[Fig.\,14 in][]{Drlica-Wagner2015}. The Galactic extinction correction is based on the \citet{SFD1998} reddening map along with the correction coefficients of \citet{Schlafly2011}. The Cetus II main sequence is prominently visible and extends from $g_\circ \approx 20.5$\,mag down to $g_\circ = 26.3$ over three magnitudes fainter than the discovery data. The limiting magnitude of the data is $g_{lim}\sim 27.5$. Unresolved background galaxies appear as plume below $g_\circ = 24.2$ and $-0.5<(g-r)_\circ < +0.6$. The red stars in the colour interval $1.0<(g-r)_\circ < 2.0$ are the population of local Milky Way M\,dwarfs.  The 50\% photometric completeness is indicated as a dashed line and reaches $g_\circ = 26.08$\,mag. 

\begin{figure}
\begin{center} 
\includegraphics[width=1.0\hsize]{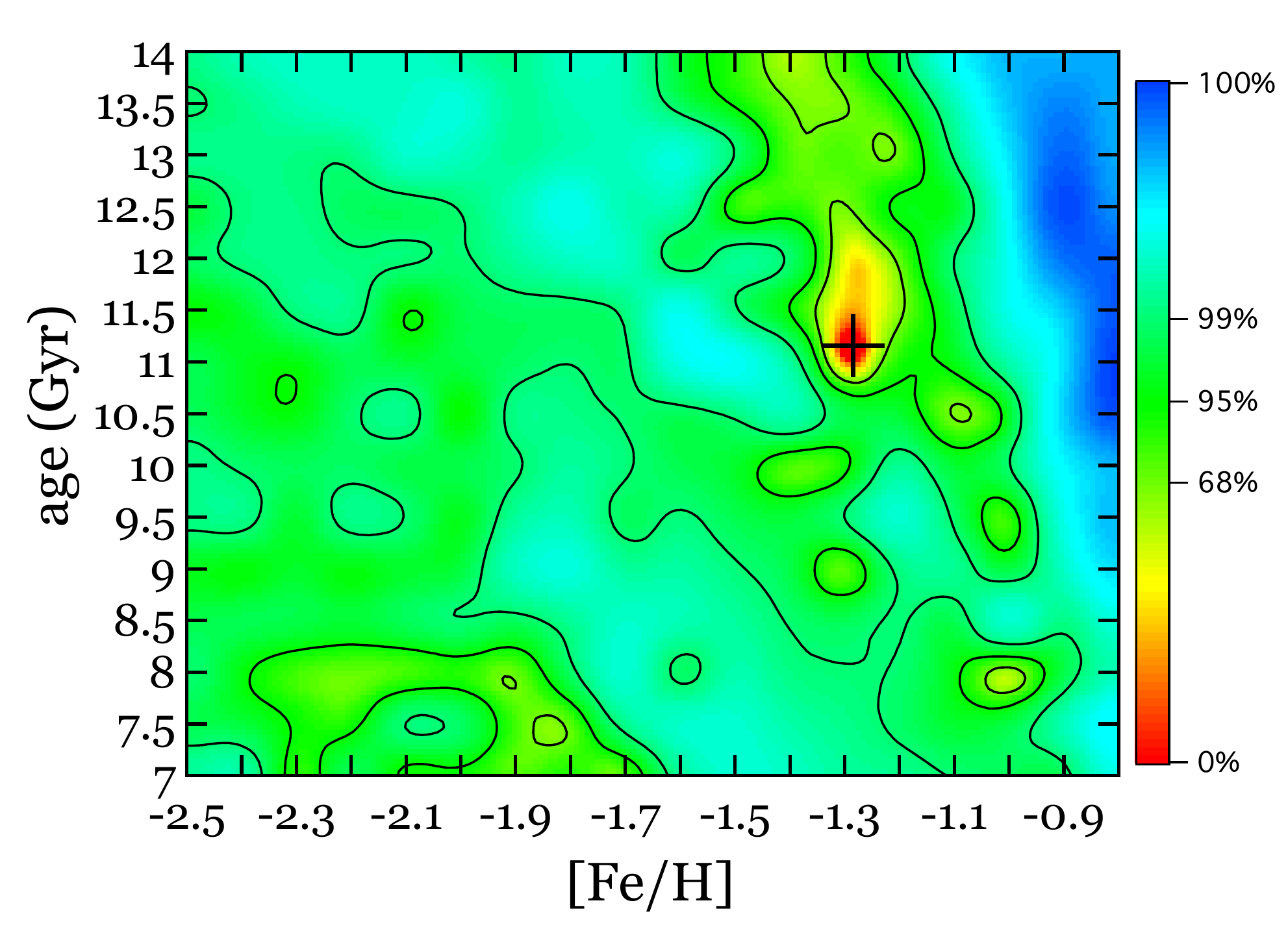}
\caption{Smoothed maximum likelihood density map in age-metallicity space for all stars within the GMOS-S field centred on Cetus II. Contour lines show the 68\% (1$\sigma$), 95\% (2$\sigma$), and 99\% (2.6$\sigma$) confidence limits. The density distribution sharply peaks for a Dartmouth model isochrone with an age of 11.2\,Gyr and a metallicity of [Fe/H]$=-1.28$\,dex. The 1D marginalized parameters around the best fit (cross) with uncertainties are listed in Table~\ref{tab:CetIIparameters}.}\label{fig:CetII_age_metal}
\end{center}
\end{figure}

\begin{figure}
\begin{center} 
\includegraphics[width=1.0\hsize]{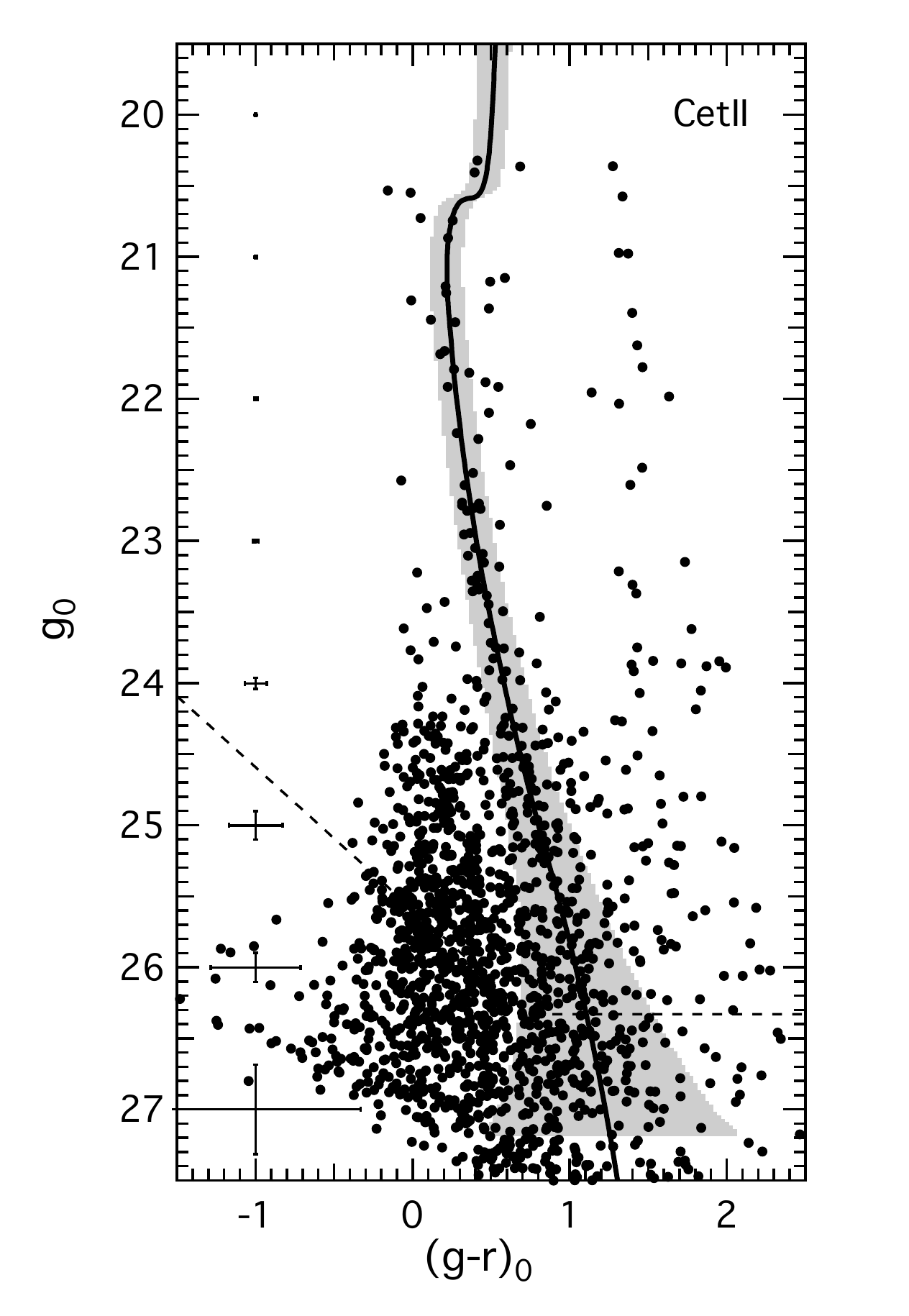}
\caption{The same colour-magnitude diagram as in Figure~\ref{fig:cmd_field} showing the best-fitting Dartmouth model isochrone and the associated mask used to identify Cetus II stars. The stellar population of Cetus II is best described by a single isochrone at a heliocentric distance of 26.3$\pm$1.2\,kpc ($m-M=17.10\pm0.10$\,mag).}\label{fig:cmd}
\end{center}
\end{figure}

\begin{figure}
\begin{center}
\includegraphics[width=1.0\hsize]{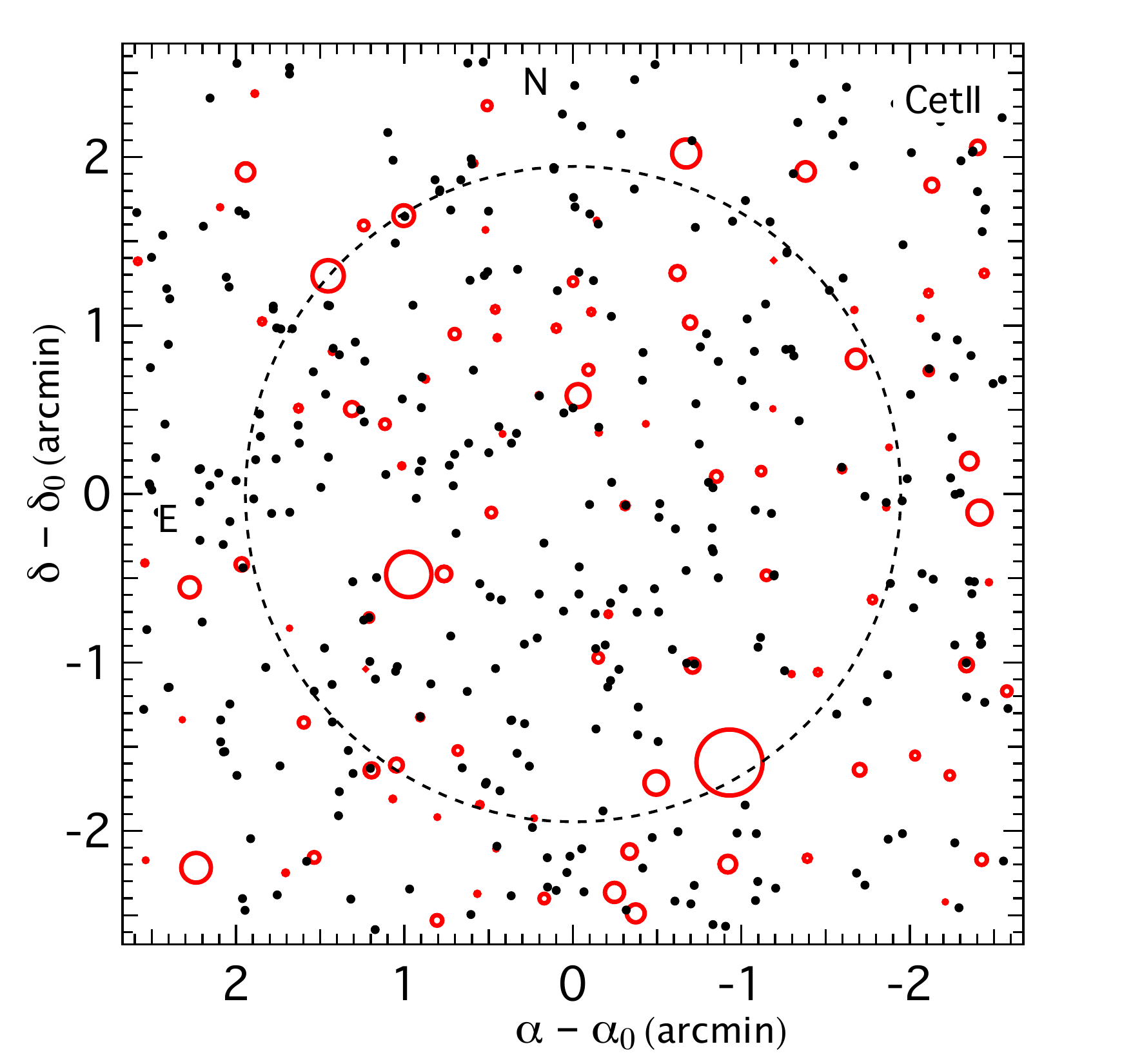}
\caption{R.A.-DEC distribution of all objects classified as stars in the GMOS-S field-of-view that pass the isochrone mask filter for Cetus II (see Figure \ref{fig:cmd}). The circle is centred on the nominal centre of Cetus II and has a radius of 1.9\,arcmin, equivalent to the reported half-light radius. Parameters were taken from \citet{ Drlica-Wagner2015}. No evidence of a stellar overdensity in the direction of Cetus II is found. Red circles are AllWISE stars \citep{Wright2010}. The diameters of the circles correlate with the brightness of the stars.}\label{fig:onskydist1}
\end{center}
\end{figure}

\section{Properties of Cetus II}\label{sec:CetIIpop}
\subsection{Stellar population}
For determining the properties of the Cetus II population we computed the model isochrone that best describes the main sequence stars distributed over the entire GMOS-S field (Figure~\ref{fig:cmd_field}) using the maximum likelihood method introduced in \cite{Frayn2002}. This method was employed in our previous studies \citep{KimJerjen2015a, Kim2, Kim2016}. In brief, we calculated the maximum-likelihood values $\mathcal{L}_i$ over a grid of Dartmouth model isochrones \citep{Dartmouth} as defined by equations\,1 and 2 in \citet{Fadely2011}. The grid points cover ages from 7--14\,Gyr, a broad range of chemical composition $-2.5\leq$ [Fe/H] $\leq-0.8$\,dex, $-0.2\leq$ [$\alpha$/Fe] $\leq +0.6$\,dex, and a distance interval $16.88<(m-M)<17.88$, where the central value of 17.38\,mag is the reported distance modulus for Cetus II from the discovery paper.  Grid steps were 0.5\,Gyr, 0.1\,dex, 0.2\,dex, and 0.05\,mag, respectively. 

Figure~\ref{fig:CetII_age_metal} shows the maximum likelihood density map of the age-metallicity space for Cetus II. The well-defined location that corresponds to the best-fitting isochrone is marked with a cross. Cetus II's stellar population is found to have an age of 11.2\,Gyr, an [Fe/H] of $-1.28$\,dex, an [$\alpha$/Fe] of 0.0\,dex and a distance modulus of $(m-M)_\circ = 17.10\pm0.10$ (26.3$\pm$1.2\,kpc). The corresponding isochrone is superimposed on the CMD in Figure~\ref{fig:cmd} together with the associated mask. The mask has an upper and lower magnitude limit of $g_o=19.5$ and 27.2, respectively. The colour width of the mask for a given magnitude $g_o$ was determined from the photometric uncertainties:
$$
(2\pi \sigma^2_{tot})^{-1/2} \exp(-((g_*-r_*)-(g-r)_{iso})^2/2\sigma_{tot}^2) >0.5,
$$
\noindent where $(g-r)_{iso}$ is the colour of the model isochrone at $g_o$ and $\sigma^2_{tot}=\sigma_{int}^2+\sigma^2_{g_*}+\sigma^2_{r_*}$. The quantity $\sigma_{int}=0.07$\,mag was chosen as the intrinsic colour width of the isochrone mask and $\sigma^2_{g_*} $, $\sigma^2_{r_*}$ are the photometric uncertainties of a star. 

We apply the isochrone mask to select the most likely Cetus II stars and plot their on-sky distribution in Figure~\ref{fig:onskydist1}. To highlight the positions of bright foreground stars in the field we also overplotted objects from the AllWISE\footnote{All Wide-field Infra-red Survey Explorer mission, http://wise2.ipac.caltech.edu/docs/release/allwise/} catalogue as red circles. The sizes of these circles correlate with the apparent magnitudes of the objects. We further show a large, dashed circle that represents the centre and half-light radius of Cetus II as reported in \citet{Drlica-Wagner2015}. Although we trace the main sequence stars over six magnitudes there is no evidence of any concentration of stars in the Cetus II field that would indicate the presence of an ultra-faint star cluster or dwarf galaxy candidate. We note that the two bright AllWISE stars in the field (see also Fig.\,2) are not affecting our ability to identify an overdensity of stars at the Cetus II location.

\begin{table}
\caption{Derived properties of the Cetus II stellar population.  \label{tab:CetIIparameters} }
{
\begin{center}
\begin{tabular}{l|c}
\hline
  & \bf{Cetus II}  \\ \hline\hline
 ($l,b$) & 
($156\fdg47,-78\fdg53$)    \\  
($\Lambda_\odot,B_\odot$) & 
($86\fdg17,8\fdg02$)    \\       
($\tilde{\Lambda}_\odot,\tilde{B}_\odot$) & ($273\fdg83,-8\fdg02$)    \\       
$E(B-V)$ & 0.0171    \\ 
$(m-M)_\circ$ & 17.10$\pm$0.10   \\
$D$ (kpc) &  26.3$\pm$1.2   \\
age (Gyr) & $11.2_{-0.4}^{+1.3}$  \\    
$[$Fe/H$]$ (dex) &  $-1.28\pm0.07$ \\   
$[\alpha$/Fe$]$ (dex)& 0.0 \\   
\hline
\end{tabular}
\end{center}
}
\end{table}

\section{Discussion}\label{sec:discussion}
As shown in $\S$\ref{sec:CetIIpop}, the Cetus II field contains a well-defined, coherent stellar population that can be traced almost six magnitudes below the main sequence turn-off. These Cetus II stars are not concentrated into a distinct stellar overdensity, and thus it was not viable to determine the centre coordinates, half-light radius, ellipticity, and total luminosity of Cetus II. Nevertheless, the properties of the underlying stellar population such as an accurate distance, age, metallicity, and alpha abundance are now accurately determined and listed in Table~\ref{tab:CetIIparameters}. We also list the Galactic coordinates, Sagittarius spherical coordinates ($\Lambda_\odot, B_\odot$) from \citet{Maj2003}, ($\tilde{\Lambda}_\odot,\tilde{B}_\odot$) from \citet{Belokurov2014} and the local dust extinction estimate from \citet{SFD1998}, based on the coordinates provided in \citet{Drlica-Wagner2015}. The Cetus II stars are old, moderately metal poor and occupy a narrow heliocentric distance range. The lack of a clear overdensity suggests these stars belong to a tidally disrupted stellar population.

\subsection{Possible connection to Sagittarius dwarf tidal stream}
\begin{figure}
\begin{center} 
\hspace{-1.0cm}
\includegraphics[width=1.1\hsize]{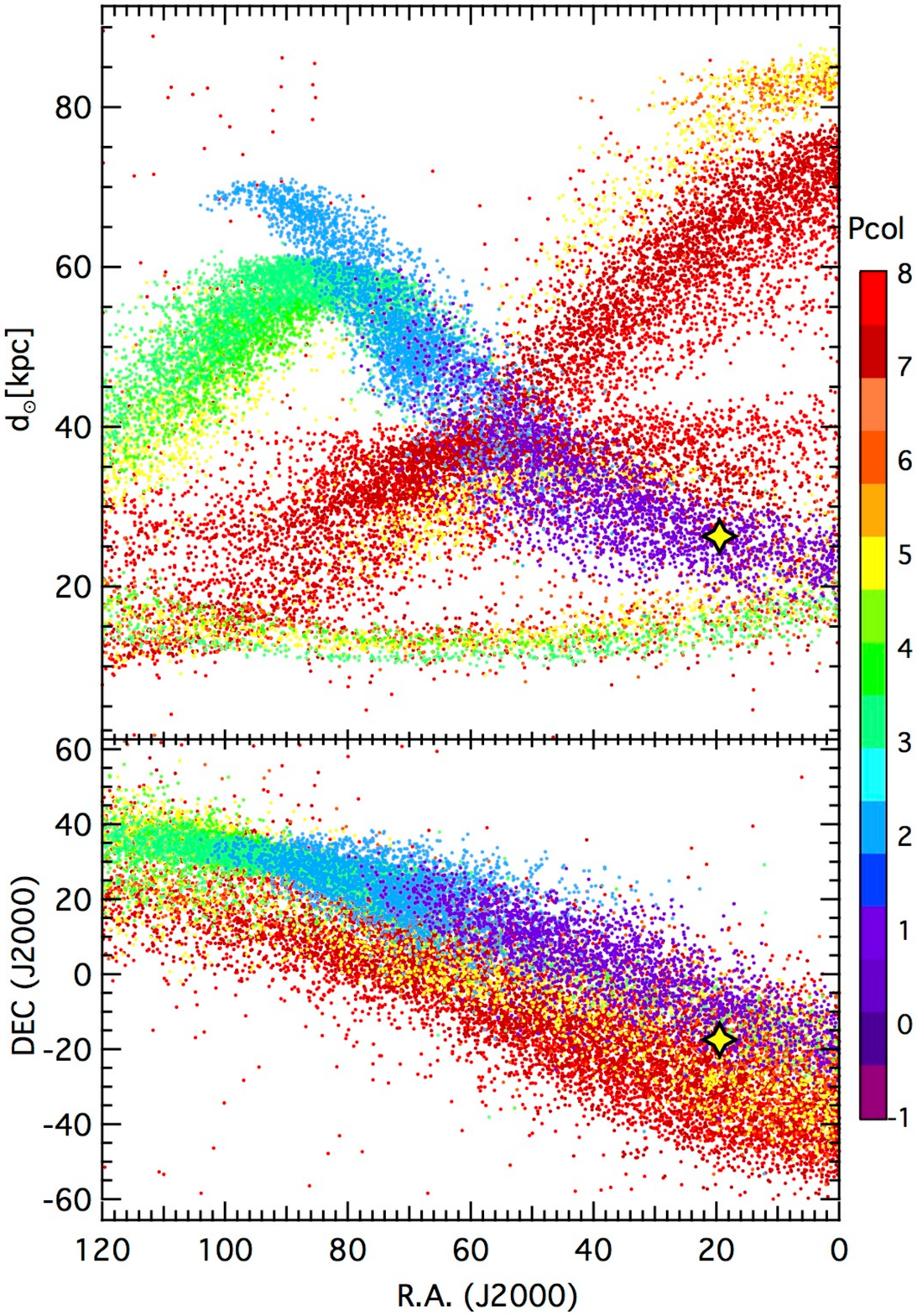}
\caption{Distribution of Sgr Stream debris particles from the \citet{LM2010} model in R.A.-distance (top) and R.A.-DEC space (bottom). Particles are colour-coded using the Pcol value. The majority of model particles in the vicinity of Cetus II have a Pcol value of 1, which indicates debris stripped on the previous pericentric passage of the Sagittarius dwarf galaxy. The position of Cetus II is indicated in both plots as yellow dot. The good agreement between model and observations supports the picture that Cetus II is a not a ultra-faint dwarf galaxy, but made up of stars in the Sgr stellar stream.}\label{fig:CSgr_LM10}
\end{center}
\end{figure}

As seen in Figure~\ref{fig:MWS}, Cetus II is projected onto the stellar density contours of the \citet{LM2010} model for the Sagittarius (Sgr) tidal stream. The presence of the Sgr stream at this location is observationally confirmed in Figure\,1 (bottom panel) from \citet{Bernard2016}. In Figure~\ref{fig:CSgr_LM10}, we use the \citet{LM2010} model to further explore the possibility that the Cetus II stars are associated with the Sgr tidal stream. In the top panel ($d_\odot$ vs R.A.), the distance and location of Cetus II is populated by model particles that were stripped on the previous pericentric passage of Sgr (Pcol=1). The \citet{LM2010} particles in a $4^\circ\times4^\circ$ window around the nominal centre of Cetus II with a Pcol value of 1 (40 particles in total) have an average heliocentric distance of 23.4\,kpc and a scatter of $\sigma = 2.9$\,kpc. Our derived Cetus II distance is in excellent agreement.
 
We further compare the Cetus II distance with the Sgr Stream distance map generated from RR Lyrae stars of type ab identified in the Pan-STARRS1 $3\pi$ survey \citep{Herni2017}. In their Figure\,1, the Cetus II stellar population is at the same distance as the Sgr stream stars cover at $(\tilde{\Lambda}_\odot,\tilde{B})\approx (274^\circ,-8^\circ)$. In Figure\,4 of the same paper, we find the Cetus II stars reside in the Sagittarius trailing arm.

\begin{figure}
\begin{center} 
\hspace{-1.0cm}
\includegraphics[width=1.1\hsize]{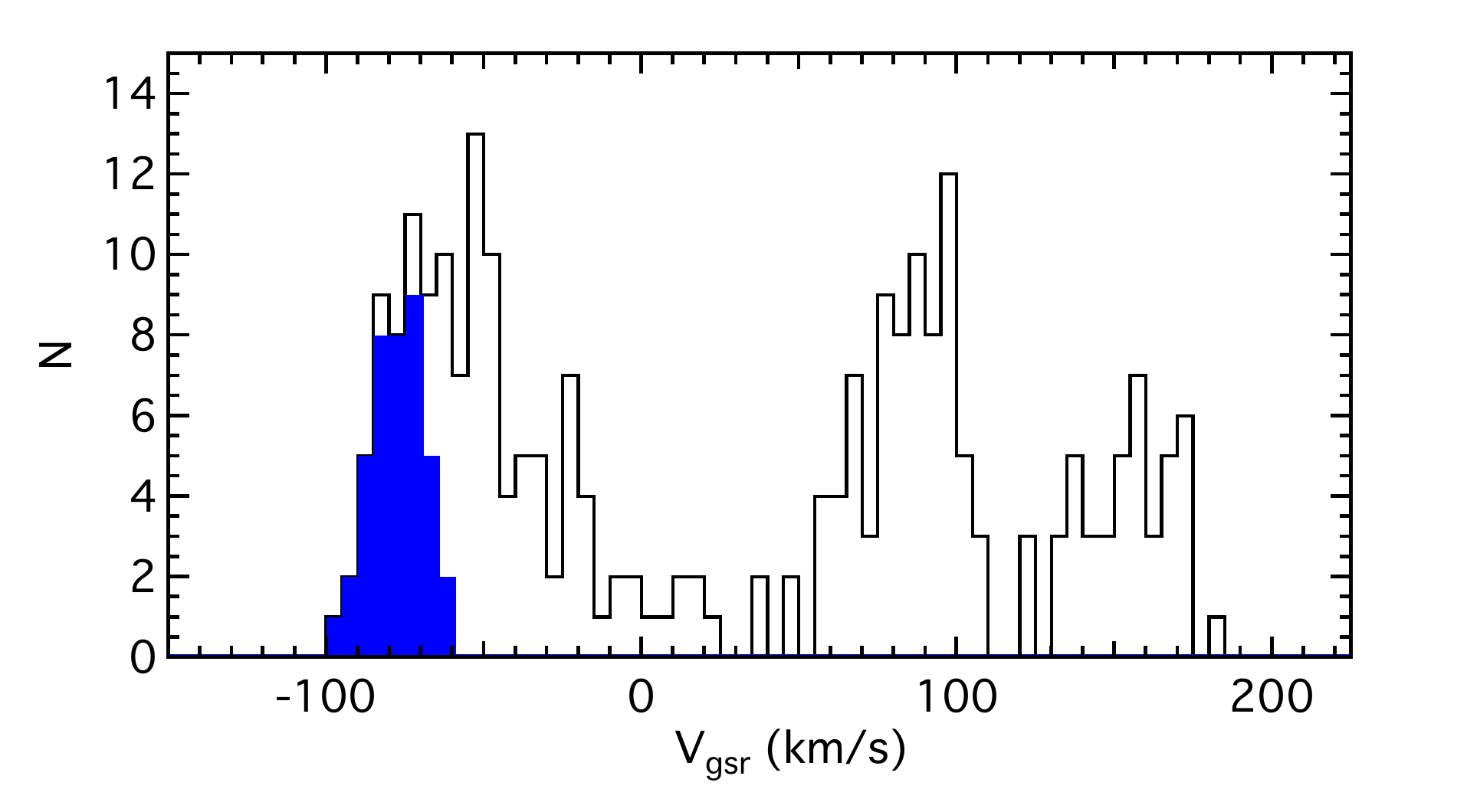}
\caption{Velocity histogram of the 245 \citet{LM2010} model particles within the $4^\circ\times4^\circ$ window centred on Cetus II. In this part of the Milky Way halo model particles cover a large range of velocities. However, the 40 particles with Pcol=1 values, model particles of the Sgr Stream that were stripped on the previous pericentric passage, have a well-defined distribution (blue histogram) with a mean of $-78$\,km\,s$^{-1}$ and a standard deviation of $9$\,km\,s$^{-1}$. Velocity measurements from spectroscopic follow-up of Cetus II stars can be used to test if Cetus II is made up of stars from that component of the trailing arm of the Sgr Stream.}\label{fig:VGSR_distr}
\end{center}
\end{figure}

In terms of age and metallicity, we also find excellent agreement between the Cetus II population ([Fe/H]$ = -1.28\pm0.07$, age$ = 11.2^{+1.3}_{-0.4}$) and the metal-poor Population\,B of Sgr  ([Fe/H]$ =-1.2\pm0.1$, age$ =11\pm1$\,Gyr), e.g. \citet{LM2010} and \citet{Siegel2007}. The final confirmation of these stars belonging to the Sagittarius tidal stream would be a spectroscopic investigation to determine their radial velocities. The \citet{LM2010} model prediction for stars in this part of the Milky Way halo is velocities in the range $-95$\,km\,s$^{-1}<v_{GSR}<-60$\,km\,s$^{-1}$ with a mean of $-78$\,km\,s$^{-1}$ and a standard deviation of $9$\,km\,s$^{-1}$, see Figure~\ref{fig:VGSR_distr}. At the location and distance of Cetus II, the Population\,B (Pcol=1) Sgr stars represent approximately 75\% of the Sgr stars in the field. The photometry presented here will be used as the basis for future spectroscopic follow-up of the Cetus II region.

\subsection{Other stellar streams}
Although we found strong evidence that Cetus II stars are part of the Sgr Stream trailing arm, we briefly want to look into other potential explanations for the Cetus II phenomenon. Is there another stream candidate which might explain the Cetus II stellar population? In their recent study of Milky Way halo substructures from the Pan-STARRS1 $3\pi$ Survey, \citet{Bernard2016} did not report a new stream nor is there any obvious candidate stream visible in their Figure\,1 at the Cetus II location. The next best alternative origin for the Cetus II stars would be the Cetus Polar Stream \citep{Newberg2009}. 

\begin{figure}
\begin{center} 
\hspace{-1.0cm}
\includegraphics[width=1.1\hsize]{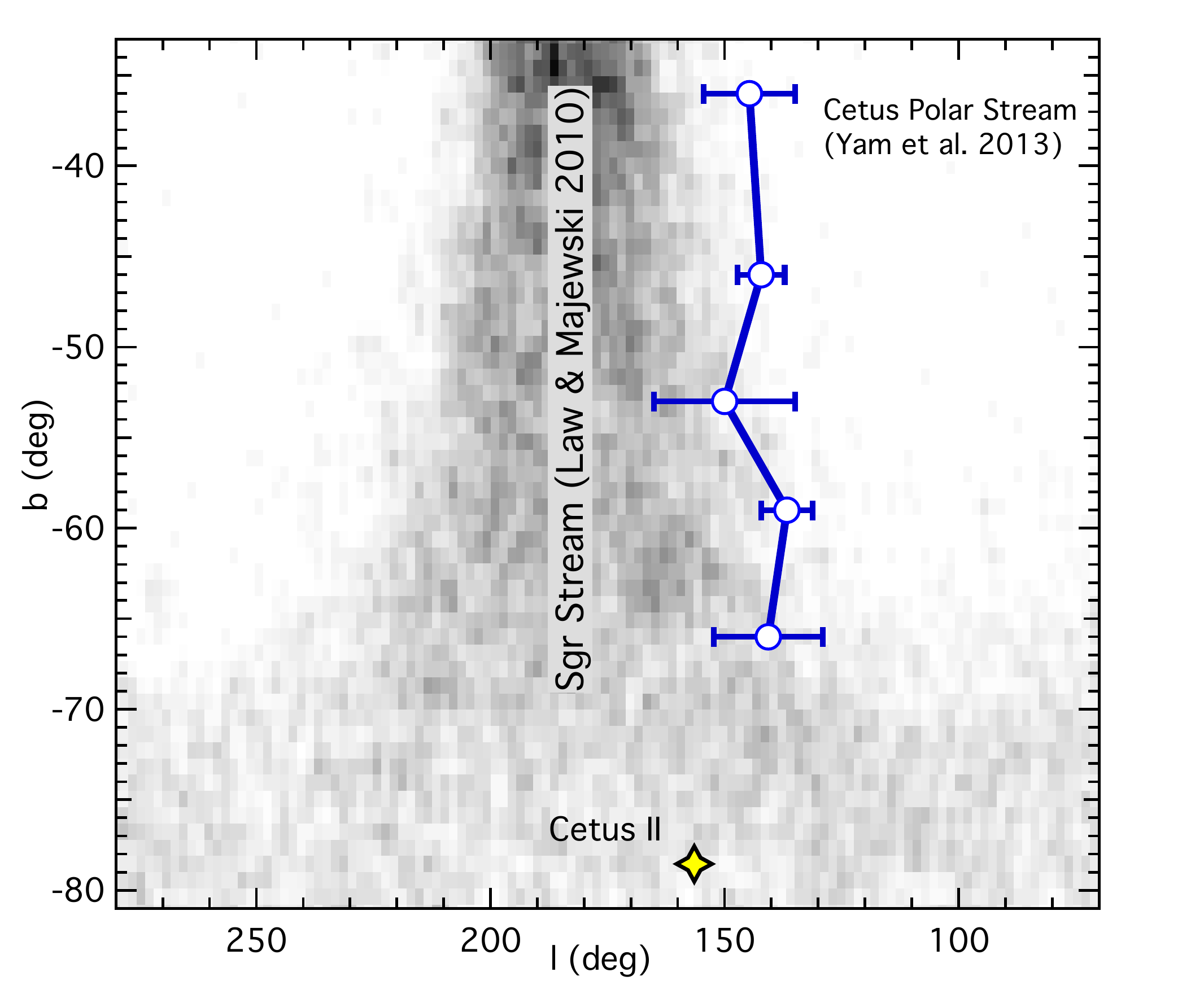}
\caption{Density distribution of particles from the Sgr Stream simulation \citep{LM2010}. Overlaid are Cetus II (yellow filled circle) and the five positions along the Cetus Polar Stream as measured from blue horizontal branch stars by \citet{Yam2013}. The width of the stream for each position is represented by a horizontal error bar. The heliocentric distance gradient of the CPS goes from 27.2\,kpc at $b\sim -36^\circ$ to 32.5\,kpc at $b\sim -66^\circ$. }\label{fig:CPS}
\end{center}
\end{figure}

In Figure \ref{fig:CPS}, we compare the density distribution of particles from the Sgr Stream simulation \citep{LM2010} with five positions along the Cetus Polar Stream (CPS) measured using blue horizontal branch stars by \citet{Yam2013}. The width of the CPS stream for each data point ($\sigma_l$ in Table\,1 of \cite{Yam2013}) is represented by a horizontal error bar. The $(l,b)^\circ$ coordinates of Cetus II disagree with the N-body simulation of a satellite on the best-fitting CPS orbit (see Fig.\,18 in \citet{Yam2013}. Moreover, the heliocentric distance of the CPS increases systematically from 27.2\,kpc at $b\sim -36^\circ$ to 32.5\,kpc at $b\sim -66^\circ$. This gradient is suggesting an extrapolated distance of $\approx 35$\,kpc at the Galactic latitude of Cetus II. Hence, the measured distance of Cetus II (26.3$\pm$1.2\,kpc) seems incompatible. Finally, the CPS is reported in \citet{Yam2013} to have a metallicity of $-2.5 < [Fe/H] <-2.0$ which is significantly more metal-poor than the stellar population in the Cetus II field.

\subsection{False-positive photometric detections}\label{sec:falsepositive}
The Cetus II case, as with Tucana V \citep{Conn2018}, highlights the complication that there are objects within the current set of candidate Milky Way satellites that are potential false-positive detections. Since they consist of a coherent stellar population but with no central overdensity these detections could be part of tidally disrupted stellar systems, which have either small physical sizes (e.g.~Kim\,1, \citet{KimJerjen2015a}) or are part of a much larger stream (e.g.~Sgr tidal Stream). The shallow photometric depth of typical discovery data combined with perhaps the apparent random clustering of stars are driving the discovery of these objects prior to a robust confirmation of their status.

The presence of a coherent stellar population combined with marginal evidence for clustering does not automatically guarantee they belong to a stellar overdensity as it was demonstrated in \citet{Jerjen2013} where multiple stellar populations were detected without being associated with a specific object or overdensity. How then are these stellar populations being identified as overdensities? Can we attribute this to solely the chance clustering of member stars or are they examples of some underlying substructure?

Before the revision of their nature, Tucana V and Cetus II curiously occupied a very similar regime in the size-luminosity plane that we have dubbed the "Trough of Uncertainty" (TUC) in \citet{Conn2018}. There were four objects in TUC of which two are now known not to be star clusters or dwarf galaxies. 

\subsection{Is spectroscopy the solution? }
Of the two remaining objects, Draco II is the brightest TUC object and has been tentatively confirmed as an ultra-faint dwarf galaxy by \citet{Martin2016a} using KECK/DEIMOS spectroscopic data, but without improving on the shallow PanSTARRS1 discovery CMD ($g_{lim}\sim22.0$, Fig.1, \citet{Laevens2015b}). DESJ0225+0304 \citep{Luque2017} is yet to be followed up. 

While spectroscopy is a good tool to test the presence of a discrete stellar overdensity, the measured velocity dispersion often used to derive the total mass is not a clean cut indicator of what type of object is being probed. For example, the classical Milky Way dwarf galaxy satellites have velocity dispersions of $\sim 10$ km\,s$^{-1}$ \citep{Walker2007} while the ultra faint dwarf satellites are in the range $3.3 - 7.6$ km\,s$^{-1}$ \citep{SG2007, Simon2015, Kirby2015}. The Milky Way globular cluster population on the other hand has velocity dispersions approximately $6\pm4$ km\,s$^{-1}$ \citep{Harris2010}. In many cases, the values are broadly consistent with the expected velocity dispersion of stars in a stellar stream, e.g.~the Sagittarius tidal stream of $\sim 9$ km\,s$^{-1}$ at the location of Cetus II. It is for this reason that confirmation and refinement of the physical properties (e.g.~half-light radius, radial profile, ellipticity and evidence of tidal disruption) of candidate ultra-faint objects are vital if we are to determine what sort of structures we are probing at these scales. However, as has been seen with Tucana V \citep{Conn2018}, scaling relations such as those presented in \citet{Forbes2014} are not applicable if, as with Cetus II, a central overdensity cannot be identified.

\subsection{Detection thresholds for targeted deep imaging}
\begin{figure}
\begin{center} 
\includegraphics[width=0.9\hsize]{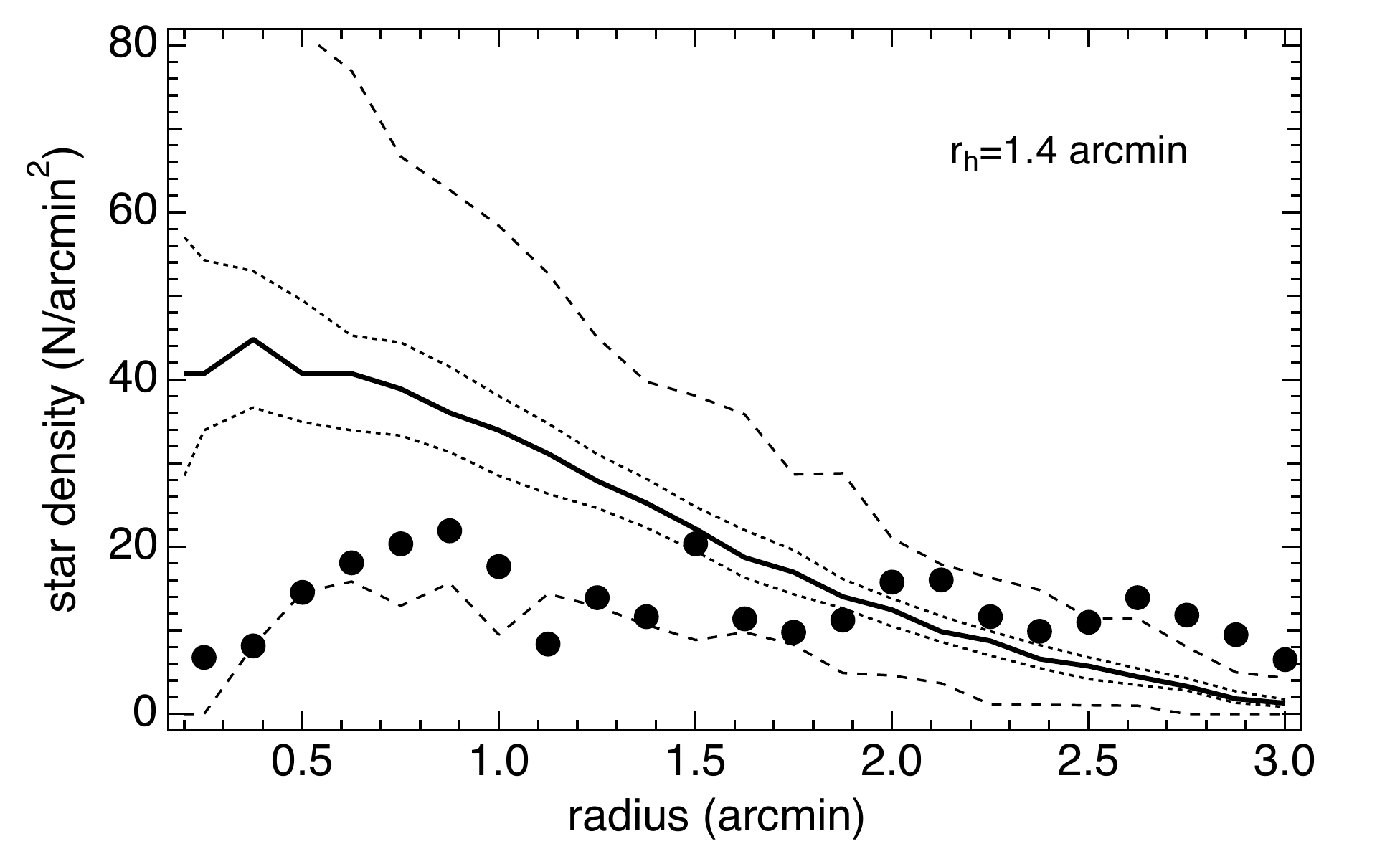}
\includegraphics[width=0.9\hsize]{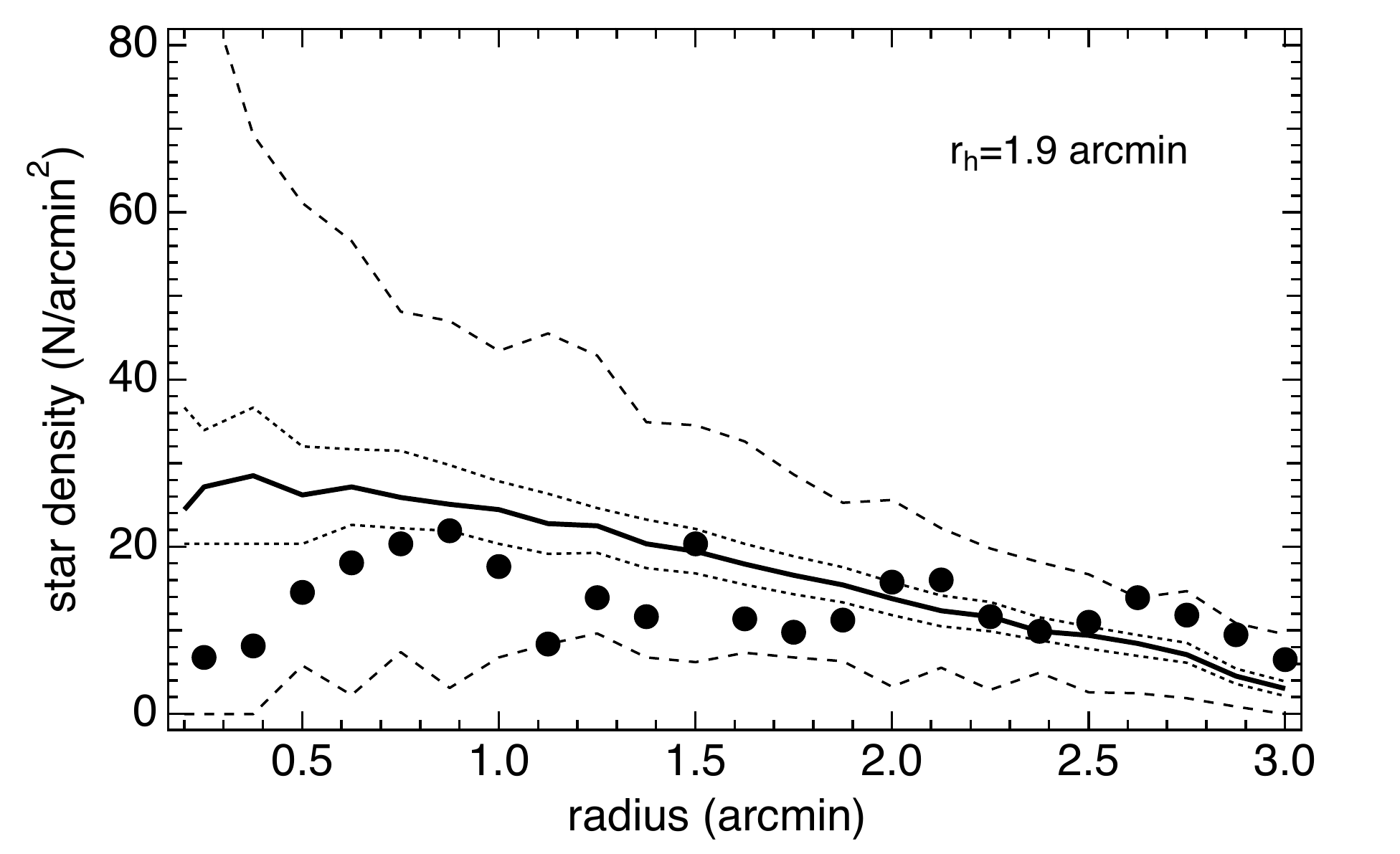}
\includegraphics[width=0.9\hsize]{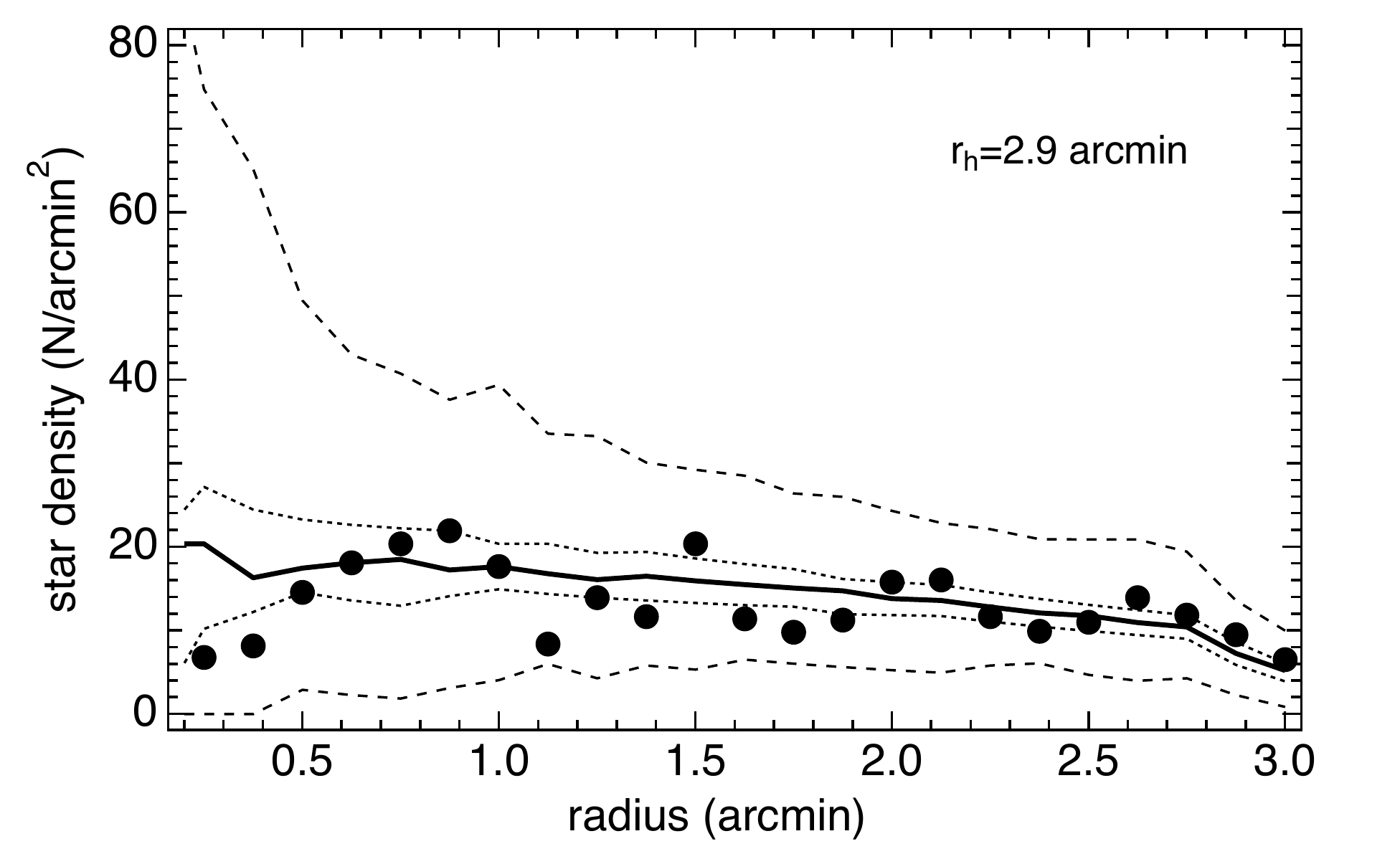}
\caption{Comparison of radial density profiles of a Cetus II-type object using parameters from \citet{Drlica-Wagner2015} with the GMOS-S data presented here. The solid lines are the median from 1000 realizations of an artificial galaxy with that particular half-light radius assuming it contains the same number of stars as seen in isochrone mask from Figure~\ref{fig:cmd}. The dotted lines are the 1st and 3rd Quartiles while the dashed lines show the minimum and maximum solutions. The solid points are the radial density profile of the stars in the Cet\,II field (Fig.\ref{fig:onskydist1})).}\label{fig:detect}
\end{center}
\end{figure}

The original discovery of Cetus II was made using DECam with its very wide field of view, however for low surface brightness systems such as these, the question then arises can deep observations on a smaller field of view adequately describe the system? In Figure~\ref{fig:detect}, we present the expected radial density profile for a Cetus II-like model galaxy assuming the parameters from \citet{Drlica-Wagner2015} and comparing them with the radial density profile as calculated using our data. In each case, the object has a 2D Gaussian profile centred on the Cetus II position published in \citet{Drlica-Wagner2015} and is populated with the 393 stars located in the Cetus II mask (see Fig.~\ref{fig:cmd}). The three panels correspond to the $1\sigma$ range of possible half-light radii reported in the discovery paper: $r_h=1.9^{+1.0}_{-0.5}$ arcmin. For each half-light radius, the stellar distribution was drawn 1000 times and a radial density profile was calculated. The lines plotted in each panel represent the properties of the total sample for a given radius. The solid line is the median, the dotted lines delimit the 1st and 3rd Quartiles and the dashed lines show the minimum and maximum values at that radius. The solid circles are the radial profile as generated using our GMOS-S data.

In the top panel of Figure~\ref{fig:detect}, the most compact of the three scenarios, with a half-light radius of 1.4 arcminutes, demonstrates that there are significant discrepancies between the data and model. Inside the half-light radius, the model consistently over-predicts the star density per radius compared to the data while at larger radii, the data outstrips the model. The middle panel, with r$_h$ = 1.9 arcminutes, the model is marginally more consistent at larger radii but the core density is still over-populated. For the largest half-light radius (r$_h$ = 2.9 arcminutes), the median radial profile is very flat, but once again the model predicts at least 20 more stars in the innermost radii than is not seen in the data. Even with such a flat profile, an extra 20 stars at the centre of the field would be easily detected in high quality data like these presented here.

\section{Conclusion}\label{sec:conclusion}
We have confirmed the presence of the Cetus II stellar population and detected member stars up to six magnitudes below the main sequence turn-off. Despite this finding, there is no overdensity in the GMOS-S field that could represent an ultra-faint star cluster or dwarf galaxy. Our photometric completeness estimates and examination of the field show that there is neither a crowding issue nor significant loss of coverage due to bright stars. Comparisons with model ultra-faint dwarf galaxies of various sizes has illustrated that even in the case with the largest half-light radius, the object should still be detectable by GMOS-S with a small field of view. It appears the original detection of an overdensity is perhaps another chance grouping of bright member stars, as is suspected with Tucana V. 

The Cetus II stars bear striking resemblance to the Population\,B stars of the Sagittarius dwarf galaxy tidal stream \citep{LM2010} in age, metallicity and distance. It is almost certain that this is the best explanation for these stars. The \citet{LM2010} models make predictions for the radial velocity distribution of these stars in the Cetus II field and spectroscopic follow-up of these stars would provide the final confirmation of their status.

Cetus II is the second object that we confirm as a false-positive detection of a stellar overdensity and it accentuates the fine line that survey teams employ when setting the criteria for their detection thresholds. This is partly due to the desire to detect the missing Milky Way satellites predicted by lambda cold dark matter (LCDM) cosmological models and partly due to the lack of a good understanding what objects like Cetus II and Tucana V might actually entail and how to classify them. 
It cannot be stressed enough that these objects consist of a single age-metallicity stellar population and thus are not misinterpretations of the colour-magnitude space. They do reside at the distances originally estimated but they do not form overdensities which conform to our understanding of star clusters or dwarf galaxies. It is most likely they represent the stochastic process of tidal disruption and as such may be providing clues regarding the structure of the object from which they were stripped. 

Given the accumulation of false-positives and the prospect of even more such detections in upcoming surveys, it is highly desirable to avoid giving ultra-faint dwarf galaxy candidates names based on the constellation they are found in, so as not to confuse the community. Candidates should keep their working names until their true nature has been unambiguously established.

\section{Acknowledgements}
BCC and HJ acknowledge the support of the Australian Research Council through Discovery project DP150100862. The authors thank the anonymous referee for their insights and improvements to this paper.

This paper is based on observations obtained at the Gemini Observatory (GS-2017B-Q-40), which is operated by the Association of Universities for Research in Astronomy, Inc., under a cooperative agreement with the NSF on behalf of the Gemini partnership: the National Science Foundation (United States), the National Research Council (Canada), CONICYT (Chile), Ministerio de Ciencia, Tecnolog\'{i}a e Innovaci\'{o}n Productiva (Argentina), and Minist\'{e}rio da Ci\^{e}ncia, Tecnologia e Inova\c{c}\~{a}o (Brazil). 

This research has made use of: the AAVSO Photometric All-Sky Survey (APASS), funded by the Robert Martin Ayers Sciences Fund;  SIMBAD database, operated at CDS, Strasbourg, France.

\software{Aladin \citep{2000A&AS..143...33B, 2014ASPC..485..277B}, Astropy \citep{Astropy2013}, DOLPHOT \citep{2000PASP..112.1383D}, TOPCAT \citep{2005ASPC..347...29T} }

This project used public archival data from the Dark Energy Survey (DES). Funding for the DES Projects has been provided by the U.S. Department of Energy, the U.S. National Science Foundation, the Ministry of Science and Education of Spain, the Science and Technology Facilities Council of the United Kingdom, the Higher Education Funding Council for England, the National Center for Supercomputing Applications at the University of Illinois at Urbana-Champaign, the Kavli Institute of Cosmological Physics at the University of Chicago, the Center for Cosmology and Astro-Particle Physics at the Ohio State University, the Mitchell Institute for Fundamental Physics and Astronomy at Texas A\&M University, Financiadora de Estudos e Projetos, Funda\c{c}\~{a}o Carlos Chagas Filho de Amparo \`{a} Pesquisa do Estado do Rio de Janeiro, Conselho Nacional de Desenvolvimento Cient\'{i}fico e Tecnol\'{o}gico and the Minist\'{e}rio da Ci\^{e}ncia, Tecnologia e Inova\c{c}\~{a}o, the Deutsche Forschungsgemeinschaft and the Collaborating Institutions in the Dark Energy Survey. The Collaborating Institutions are Argonne National Laboratory, the University of California at Santa Cruz, the University of Cambridge, Centro de Investigaciones En\'{e}rgeticas, Medioambientales y Tecnol\'{o}gicas-Madrid, the University of Chicago, University College London, the DES-Brazil Consortium, the University of Edinburgh, the Eidgen\"{o}ssische Technische Hochschule (ETH) Z\"{u}rich, Fermi National Accelerator Laboratory, the University of Illinois at Urbana-Champaign, the Institut de Ci\`{e}ncies de l'Espai (IEEC/CSIC), the Institut de F\'{i}sica d'Altes Energies, Lawrence Berkeley National Laboratory, the Ludwig-Maximilians Universit\"{a}t M\"{u}nchen and the associated Excellence Cluster Universe, the University of Michigan, the National Optical Astronomy Observatory, the University of Nottingham, the Ohio State University, the University of Pennsylvania, the University of Portsmouth, SLAC National Accelerator Laboratory, Stanford University, the University of Sussex, and Texas A\&M University.

\end{document}